\documentclass[iop,apj]{emulateapj}

\usepackage{multirow}
\usepackage{hhline}

\newcommand\Tstrut{\rule{0pt}{2.6ex}}       
\newcommand\Bstrut{\rule[-0.9ex]{0pt}{0pt}} 
\newcommand{\TBstrut}{\Tstrut\Bstrut} 

\slugcomment{ApJ}

\shorttitle{Type Ia Supernova Rates and Progenitors}
\shortauthors{Heringer et al.}

\begin{document}


\title{Type Ia Supernovae: Colors, Rates, and Progenitors}


\author{Epson Heringer\altaffilmark{1,2}, 
Chris Pritchet\altaffilmark{1}, 
Jason Kezwer\altaffilmark{1},
Melissa L. Graham\altaffilmark{3},
David Sand\altaffilmark{4},
Chris Bildfell\altaffilmark{1}
}


\altaffiltext{1}{Department of Physics and Astronomy, University of Victoria, PO Box 1700 STN CSC Victoria, BC  V8W 2Y2}
\altaffiltext{2}{Department of Astronomy \& Astrophysics, University of Toronto, 50 St. George Street, Toronto, ON, M5S 3H4}
\altaffiltext{3}{Department of Astronomy, University of Washington, Box 351580, U.W., Seattle WA 98195-1580}
\altaffiltext{4}{Texas Tech University, Physics Department, Box 41051, Lubbock, TX 79409-1051, USA}


\begin{abstract}

The rate of type Ia supernovae (SNe\ Ia) in a galaxy depends not only on stellar mass, but also on star formation history. Here we show that two simple observational quantities ($g-r$ or $u-r$ host galaxy color, and $r$-band luminosity), coupled with an assumed delay time distribution (the rate of SNe\ Ia as a function of time for an instantaneous burst of star formation), are sufficient to accurately determine a galaxy's SN\ Ia rate, with very little sensitivity to the precise details of the star formation history. Using this result, we compare observed and predicted color distributions of SN\ Ia hosts for the MENeaCS cluster supernova survey, and for the SDSS Stripe 82 supernova survey. The observations are consistent with a continuous delay time distribution (DTD), without any cutoff. For old progenitor systems the power-law slope for the DTD is found to be $-1.50 ^{+0.19} _{-0.15}$.  This result favours the double degenerate scenario for SN\ Ia, though other interpretations are possible. We find that the late-time slopes of the delay time distribution are different at the 1$\sigma$ level for low and high stretch supernova, which suggest a single degenerate scenario for the latter. However, due to ambiguity in the current models' DTD predictions, single degenerate progenitors can neither be confirmed as causing high stretch supernovae nor ruled out from contributing to the overall sample.

\end{abstract}





\def\spose#1{\hbox to 0pt{#1\hss}}
\def\lta{\mathrel{\spose{\lower 3pt\hbox{$\mathchar"218$}}
     \raise 2.0pt\hbox{$\mathchar"13C$}}}
\def\gta{\mathrel{\spose{\lower 3pt\hbox{$\mathchar"218$}}
     \raise 2.0pt\hbox{$\mathchar"13E$}}}
\def\up#1{{$^{#1}$}}
\def\addrefs{{\bf addrefs~}}
\def\addref{{\bf addref~}}

\section{Introduction}
\label{sec:intro}

\citet{Hoyle_intro1} were the first to suggest that at least some
supernovae originate in explosions of degenerate C+O white dwarfs;
there is strong circumstantial \citep{Pritchet2008_dtd, Bloom2012_WD}
and direct \citep{Bloom2012_WD, Nugent2011_WD, Piro_intro5} evidence
that this picture is substantially correct. We now know that roughly 1\%
of all white dwarfs (WDs) eventually end their lives as
Type Ia supernovae (SNe\ Ia; \citealt{Pritchet2008_dtd}), and that
these WDs are almost certainly members of binary or multiple star
systems; yet the precise progenitor and mechanism that leads to an 
explosion remains elusive. 


The importance of understanding SN\ Ia progenitors and explosion
mechanisms cannot be overemphasized. Dark energy was discovered using
observations of intermediate redshift ($z < 1$) SNe\ Ia \citep{Riess1998_de, Perlmutter1999_de}; more recent observations of larger samples of
supernovae have been used to constrain the nature of dark energy
\citep{Sullivan2011_SNLS}. However, only with a better understanding of the
supernova explosions themselves can we hope to understand the
calibration of SNe\ Ia as distance indicators, and evaluate the effects
of potential redshift-dependent systematics. Furthermore, SNe\ Ia are the 
primary source of Fe-peak elements; understanding the origin of SNe\ Ia is therefore a
key to mapping the buildup of metals in the early Universe
(e.g. \citealt{Thielemann1986_chem, Tsujimoto1995_chem}).

Most of the discussion of the origin of SNe\ Ia centers on the single degenerate (SD) and double degenerate (DD) scenarios.
In the SD scenario \citep{Whelan1973_SD, Nomoto2000_SD_MSandRG}, a Type Ia supernova occurs when a CO WD accretes enough material from an
evolving binary companion to drive it close to the Chandrasekhar
limit, $M_{\mathrm{Ch}}\simeq 1.4 $M$_\odot$ \citep{Chandrasekhar1931_mass}. A
merger of two WDs (the DD scenario) may also lead
to a SN\ Ia progenitor if the combined mass exceeds $M_{\mathrm{Ch}}$
\citep{Tutokov1981_DD, Webbink1984_DD}. In addition there exist sub-Chandrasekhar mass models
(involving double detonations; e.g. \citealt{Shen2010_doubledet}) and
core degenerate models (involving the merger of a WD with the core of
an AGB star; e.g. \citealt{Soker_intro11}). \citet{Kerkwijk2010_dtd}
proposes that the DD channel can trigger SNe\ Ia even if the
combined mass of the WDs does not exceed $M_{\mathrm{Ch}}$. Many problems exist
in the explosion physics and theoretical rates for these and other
scenarios \citep{Wang2012_dtd, Hillebrandt_intro12}.

SNe\ Ia were originally assumed to belong exclusively to
old stellar populations, due to their occurrence in elliptical galaxies. \citet{Branch1993_history} were among the first to show that SNe\ Ia also occur in young stellar populations; it is now known that SNe\ Ia are a factor $\sim
10-30$ more frequent (per unit mass) in young starburst galaxies than
in old galaxies \citep{Sullivan2006_SNrates, Smith2012_RSFcorrelation}. This is not a surprising result, given that, through mass loss, stars as massive as $\sim 5.5$ M$_\odot$ (main sequence lifetimes $\sim 100$ Myr) can end their lives as CO WDs \citep{Chen2014_mass}. (Stars with initial masses $5.5 \lta $ M/M$_\odot$ $\lta 8$ end their lives as O+Ne WDs, which probably do not explode (\citealt{Chen2014_mass}, though see \citealt{Marquardt2015_ONe}). Stars more massive than $\sim$ 8 M$_{\odot}$ end their lives as either black holes or neutron stars after a core collapse supernova \citep{Heger2003_WDmass}.) 

What is known about the rates of SNe\ Ia? The SN\ Ia delay time distribution (DTD) gives the rate of SNe\ Ia from an instantaneous burst of star formation as a function of time, normalized to the burst mass. (The stellar mass of a burst decreases with time due to evolution and mass loss. By convention the DTD is normalized to the {\it initial} mass formed in a burst of stars.) The DTD depends on, and provides insight into, the progenitor channel. Unfortunately, binary population synthesis models (e.g. \citealt{Maoz2014_dtd} and references therein) give widely disparate estimates of the DTD (as much as a factor of $10^3$ different in some cases), and underestimate the overall rates by as much as a factor of $\sim 10$ (e.g. \citealt{Wang2012_dtd}).

The observational situation is somewhat more encouraging: the
 DTD has been found to vary roughly as $t^{-1}$
\citep{Totani2008_powerlaw, Maoz2010_dtd, Graur2011_dtd, Maoz2012_fig,
MaozandMannucci2012_dtd, Sand2012_MENeaCS_SNsurvey, Graur2013_fig}, a functional form that is appealing because it agrees with
simple gravitational energy loss timescales for merging WDs
\citep{Maoz2014_dtd}. But determining the DTD 
relies on measuring the ages of individual galaxies, and these ages
are notoriously inaccurate, especially in the presence of
complex star formation histories. Even more troublesome is the fact that
SN progenitors are drawn with non-uniform probability from
the age mix of stars in a galaxy; in other words the mean stellar age
may not necessarily be the mean age of SN\ Ia progenitors. As a simple toy model,
consider an old galaxy with a very weak
burst of recent star formation, and a DTD with a cutoff at late times (i.e. an upper
limit on the age of progenitors). SNe\ Ia will occur in the young stars
in this galaxy, but these supernovae may be
(erroneously) attributed to the dominant old population.
In more detail, if DTD $\propto t^{-1}$, it follows that the rate of SNe\ Ia per unit
mass can vary by a factor of $100$ depending on the age of a
star burst (which can vary from roughly $10^8$ to 10$^{10}$ yr). A
$1\%$ (by mass) burst of star formation can in principle provide a
SNe\ Ia rate equivalent to that of the (much more massive) old population
of a galaxy. Is it therefore possible that SNe\ Ia in old elliptical
galaxies originate {\it only in} a ``frosting" of young stars, and that all
SNe\ Ia belong to a relatively young ($\lta 10^9$ yr old) stellar population? 
Put a different way, is there a cutoff in the SN\ Ia DTD?

There is in fact a physical basis for a DTD cutoff at late times.
Primary stars with a mass less than 2 M$_\odot$ (lifetime $\gta 10^9$
yr) are more likely to produce a He WD than a CO WD
\citep{Yungelson2005_HeWD}; if accretion then pushes such a star above
$M_{\mathrm{Ch}}$, it will undergo a He flash rather than a SN\ Ia explosion
\citep{Greggio2005_dtd}. Even if a CO white dwarf is produced, an
accretion rate around $10^{-7}$ M$_\odot$ yr$^{-1}$ is required for
stable burning and mass growth \citep{Nomoto1982_rate}, and this will
only occur if the secondary also has a mass $\gta 2$ M$_\odot$
\citep{Han_intro9, Maoz2014_dtd}. Thus a cutoff $t_c$ in the DTD around $10^9$ yr
might be expected for the SD scenario. (On the other hand, the
discovery of high mass-ratio binaries among massive stars
\citep{Moe2015_SD} indicates the presence of low mass main sequence
(MS) companions, potentially allowing longer delay times for the SD
scenario. This however is only relevant if the $M > 2 M_\odot$ constraint 
for the secondary is relaxed.)

The observational evidence for or against a break in the DTD is
ambiguous.  Some (but not all) of the data in the compilation of
\citet{Maoz2014_dtd} hints at a break (e.g. \citealt{Maoz2011_fig, Maoz2012_fig}), but with large uncertainties.
\citet{Schawinski2009_sn} uses GALEX $NUV-r$ color to show that SNe\
Ia observed in early-type galaxies might be due to episodes of recent
star formation, and hence that the DTD may have a cutoff.

In this paper we study the nature of the DTD and whether it
has a break or cutoff, using existing samples of SNe\ Ia. We use a simple
stellar population model to calculate supernova rates as a function of
galaxy color, for a variety of DTDs. We then predict the color distribution of SN hosts using a control sample of galaxies, which consists of galaxies from which the SN hosts were drawn. Following this, we
compare this color distribution with that observed for actual SN hosts.
In particular, we search for a deficit of SNe\ Ia at 
the color of the red sequence relative to predicted rates. If observed,
such a deficit could be interpreted as due to the effects of
a cutoff in the DTD.

\S 2 discusses the supernova surveys and control samples that were
used; \S\S 3--5 explain the stellar population models and the
methodology. \S 6 discusses the results. We use cosmological
parameters from the Planck collaboration
\citep{Planck2014_cosmologicalpars}: $H_0=67$ [km s$^{-1}$
Mpc$^{-1}$], $\Omega_\Lambda=0.68$ and $\Omega_m=0.32$.

\section{Sample selection}

We analyze two low redshift samples of supernovae and field galaxies: the Multi-Epoch Nearby Cluster Survey (MENeaCS) in $g-r$; and the Sloan Digital Sky Survey Stripe 82 (SDSS-II) in $g-r$ and $u-r$. We compute the cumulative supernova rate we would expect to observe in each survey. In order to do this, we define a control sample, which consists of galaxies that satisfy the selection criteria discussed below. 

\subsection{MENeaCS Supernova Survey}


The MENeaCS survey \citep{Sand2012_MENeaCS_SNsurvey} sampled 57 X-ray selected rich clusters with redshifts $0.05 < z < 0.15$. Repeated $g$- and $r$-band observations of these clusters were obtained over a 2 year period using the Canada-France-Hawaii Telescope with its MegaCam imager \citep{Boulade2003_megacam}. The detection limit was $g$=$r$=23.5 in the difference imaging. MENeaCS spectroscopically confirmed 23 cluster SNe\ Ia, 4 of which were almost certainly intracluster events which are not used in our analysis \citep{Sand2011_MENeaCS_survey, Graham2015_MENeaCS}. Other than for SN\ Ia hosts, some clusters have spectroscopy available in archival sources \citep{Sifon2015_MENeaCS}, including the SDSS, which we use to study environmental effects on our samples in \S \ref{sec:rsfit}. Additional $u$- and $i$-band photometry is also available (see \citealt{Burg2015_MENeaCS}), which we do not use since our results with the SDSS data show that the $g-r$ color leads to smaller uncertainties.



The initial control sample contains 57,638 galaxies. For simplicity, we adopt a redshift $z=0.1$ for all objects to compute the absolute magnitude in the $r$-band, $M_r$. In order to exclude objects with spurious photometry we require $-23.5 \leq M_r < -17.5$, which reduces the number of galaxies in the control sample to 56,523. Out of the 19 cluster hosts, five do not satisfy the color (see \S \ref{sec:methodology}) and magnitude selection criteria we adopt for our control sample; we do not consider these hosts in our analysis. Our MENeaCS sample of SN\ Ia hosts therefore contains 14 galaxies.

\subsection{SDSS-II Supernova Survey}

The SDSS-II Supernova Survey \citep{Frieman2008_SDSS_SNsurvey} covered a 300 deg$^2$ region of Stripe 82 ($-60^{\circ} < \alpha < 60^{\circ}$, $-1^{\circ}.25 < \delta < 1^{\circ}.25$) over nine months during 2005-2007 (with some engineering time in 2004). Of the more than 500 SNe\ Ia that were spectroscopically confirmed, we select those with hosts bright enough to have spectroscopy in the SDSS DR7 catalog\footnote{http://www.sdss.org/dr7/} ($14 < r < 17.77$, after correction for Milky Way extinction). We follow the host-matching procedure of \citet{Sullivan2006_SNrates}, described in detail by \citet{Gao2012_SNmatching}\footnote{See \citet{Sand2011_MENeaCS_survey} for the correction of a typo in  \citet{Sullivan2006_SNrates}.}. The final sample of SNe\ Ia is 53 in the redshift range {$0.01 < z < 0.2$}. The control sample contains 20,707 galaxies.

The analysis of stellar populations in field galaxies and SN\ Ia host galaxies uses $ugriz$ Petrosian magnitudes from the SDSS DR7 database, specifically $u-r$ and $g-r$ colors, which are the most sensitive to age and star formation effects  (in the observed $z$ range). 

\section{Color and Absolute Magnitude Calculation}

Colors and absolute magnitudes are required for the analysis we perform in this work. These are calculated as follows.

\subsection{Absolute Magnitudes}

First we correct the observed apparent magnitudes for  galactic extinction using the \citet{schlegel1998_extinctionmap} extinction map. Absolute magnitude is then calculated using:

\begin{equation}
M_{X} = m_{X} -5 log_{10} D_{L} - 25 - K_{X}(z) + Q \cdot z,
\label{eq:abs_mag}
\end{equation}

\noindent
where $X$ is passband and $D_L$ is luminosity distance. The $k$-correction to redshift zero, $K_{X}(z)$, is calculated using KCORRECT \citep{blanton2007_kcorrect}; the evolutionary correction, $Q \cdot z$, is made using $Q=1.6$ \citep{Wyder2007_cmd}.

\subsection{Red-Sequence Fitting}
\label {sec:rsfit}

The red sequence (RS) is, roughly speaking, the locus of the oldest galaxies in a color-magnitude diagram \citep{Strateva2001_RS, Wyder2007_cmd}. The color of a RS galaxy is correlated with its absolute magnitude or mass; this dependence is known as the color-magnitude relation (CMR), where the reddest of the RS galaxies are the brightest and largest. The CMR (and scatter around it) is the product of a variety of effects, but particularly variations in metallicity and age \citep{Graves2009_RS}. 

For simplicity we measure all colors relative to the fitted RS:

\begin{equation}
\Delta \equiv \Delta(g-r) = (g-r) - (g-r)_{\mathrm{RS}},
\label{eq:delta}
\end{equation}

\noindent
and similarly for other passbands. This step removes some calibration errors, extinction errors, and also the slope of the CMR if $(g-r)_{\mathrm{RS}}$ is measured at the same absolute magnitude as the galaxy in question. (As will be seen in \S \ref{sec:fspstests}, using color relative to the RS in the stellar population models has a similar beneficial effect.)

For MENeaCS clusters, each cluster RS CMR was fitted; the CMR was subtracted from the data and the clusters co-added. This procedure eliminates field-to-field calibration and extinction differences.

Although the RS is generally associated with rich clusters, a strong RS is also present for (predominantly field) galaxies in SDSS Stripe 82. To fit the CMR of this sample, we first excluded objects that were clearly not part of the RS, and then used an iterative outlier rejection method. 
The RS slope, intercept and standard deviation are given in Table \ref{tb:RS}. In particular, the fitted slope of the SDSS $g-r$ sample, $-0.019$, is in agreement with the average value found for the MENeaCS sample, $-0.026$, and with the value obtained by \citet{Hogg2004_grslope}, $-0.022$, for field galaxies in different environments. Similarly, the linear coefficient of the $u-r$ SDSS sample, $-0.069$, is in agreement with the value found by \citet{Baldry2004_urslope} for SDSS field galaxies, $-0.08$. The formal errors in these slopes are very small ($3\times 10^{-4}$ for $g-r$ and $\sim 2\times 10^{-3}$ for $u-r$), but the true uncertainties are dominated by systematic errors such as the age and metallicity dispersion in the RS.

\begin{table}[ph]
\begin{center}
\caption{$RS$ parameters. \label{tb:RS}}
\begin{tabular}{cccc}
\tableline\tableline
Color & Sample & Slope & $\sigma$ \TBstrut \\ 
\tableline\tableline
\multirow{2}{*}{$(u-r)$} & MENeaCS & - & - \Tstrut \\
& SDSS & $-0.069$  & $0.21$ \Bstrut \\
\tableline
\multirow{2}{*}{$(g-r)$} & MENeaCS & $-0.026$ & $0.039$ \Tstrut \\
& SDSS & $-0.0188$ & $0.04$ \Bstrut \\
\tableline
\end{tabular}
\end{center}
\end{table}

Is the age and metallicity of the field RS in SDSS Stripe 82 the same as the cluster RS in the MENeaCS data? To answer this question we examined regions of high density in Stripe 82 (which appear as ``stripes" in the color-redshift diagram). No difference in RS color was observed between these regions and the entire sample. We also looked at $\sim 25$ MENeaCS clusters lying in the full SDSS DR7 footprint, comparing the colors of their RS with the surrounding field. The cluster RS was found to be 0.02 (0.06) mag redder in $g-r$ ($u-r$) compared with the field, corresponding to an age difference of 1.7 (1.4) Gyr, or a metallicity difference of 0.003 (0.002). (A similar result has been found by \citealt{Hogg2004_grslope}.)  As we show later, such a small difference in the age or metallicity of the oldest populations does not affect our results.

\section{Stellar Population Models}

Galaxy colors were modelled using the Flexible Stellar Population Synthesis code (FSPS -- \citealt{Conroy2009_fspsI, Conroy2010_fspsII, Conroy2010_fspsIII}). We used the BaSeL spectral library \citep{Lejeune1997_BaSelI, Lejeune1998_BaSelII, Westera2002_BaSelIII}, combined with PADOVA isochrones \citep{Marigo2007_Padova, Marigo2008_Padova} and the Chabrier \citep{Chabrier2003_imf} initial mass function (IMF). We assumed solar metallicity, and no shift in log $L$ and log $T$ for TP-AGB stars; the dust parameters, fraction of blue stragglers (BS), and fraction of extended horizontal branch (EHB) stars were all set to zero. In \S \ref{sec:fspstests} we address the consequences of varying the metallicity and EHB fraction adopted, finding no significant impact.  

Initially we modelled the star formation history (SFH) of galaxies as a sum of two simple stellar populations (SSPs). This model has previously been employed to probe recent star formation (RSF) in early-type galaxies (e.g. \citealt{Ferreras1999_young, Ferreras2000_2B, Schawinski2007_cmd, Kaviraj2007_2B}). For this simple case there are three parameters that determine the properties of a galaxy: the ages of the young and old populations ($t_y$ and $t_o$), and the mass fraction $\alpha = M_y / M_o$ of the burst. (Here $\alpha$ is defined in terms of the stellar mass initially formed in each component, and does not include the effects of later stellar evolution and mass loss.) 

We also considered a wide range of more complicated star formation histories. Simplest of these are star formation rates that decay exponentially, SFR $\propto$ exp$(-t/\tau)$. Other models involving gas infall, star formation, and winds \citep{LeBorgne2002_pegase} were also used; for these models we tried a wide range of parameters that satisfy observed galaxy colors, with ages $\le 10$ Gyr. In all of these cases we also considered the effects of adding a burst (SSP) component with variable age and mass fraction $\alpha$.

\subsection{Age--Color Relation}

\begin{figure}
\epsscale{1.2}
\plotone{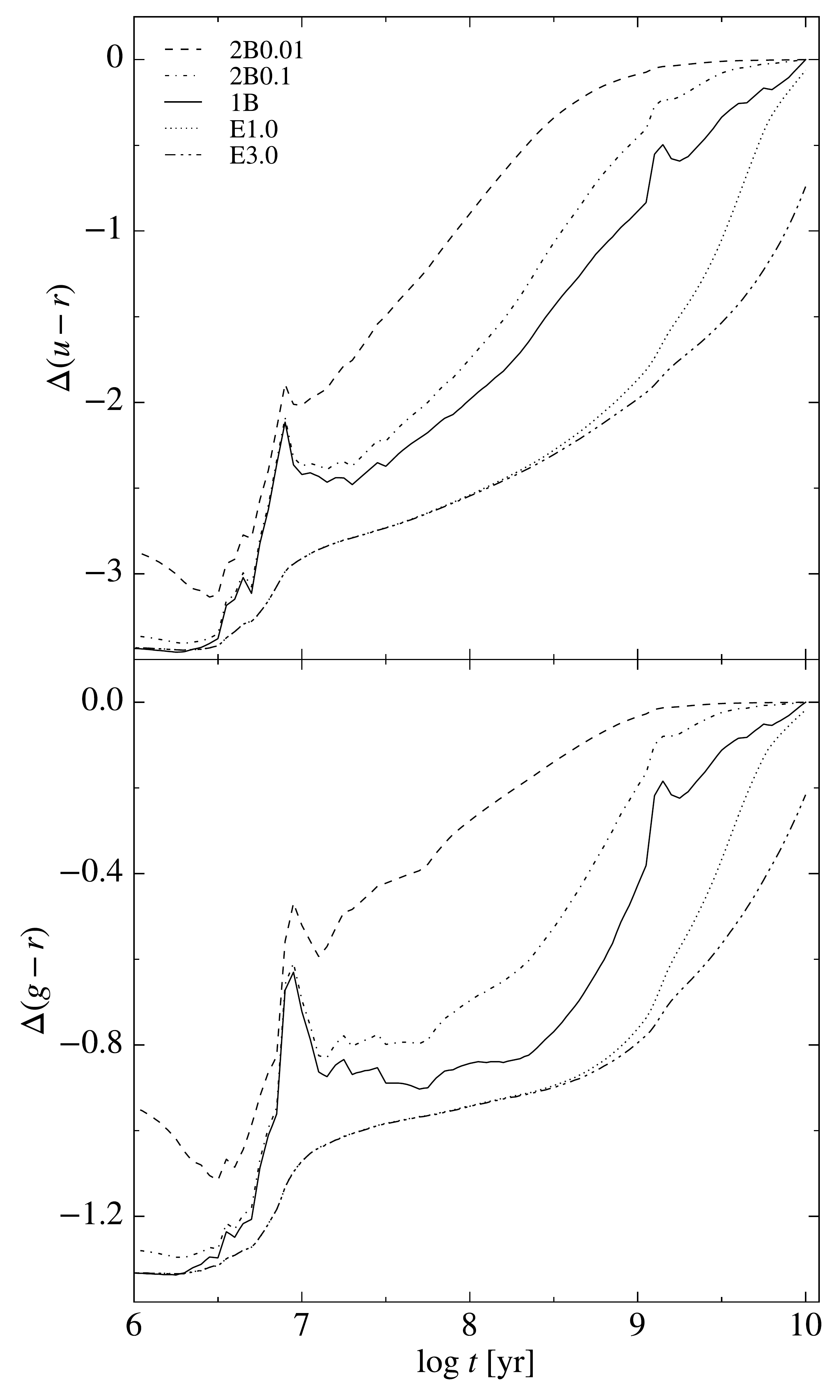}
\caption{Age versus color shift for $u-r$ and $g-r$ for a range of SFHs. The color shift is relative to the RS, which in these plots is assumed to be the color of a 10 Gyr SSP. The curves labelled 2B are double burst models (``2B0.1" indicates a mass fraction $\alpha=0.1$), and for these cases the age is the age of the {\it young} population. The solid curve labelled 1B is a SSP, i.e. $\alpha \rightarrow \infty$. Curves labelled E are exponential SFHs (``E1.0" corresponds to $\tau=1$ Gyr).}
\label{Fig:logagey_Dcolour}
\end{figure}

Using the FSPS predictions for a single SSP, we can calculate the color deviation from the RS for galaxies composed of two populations. The color of the RS is assumed to be the color of an SSP at $t_o = 10$ Gyr.

Variations of a few Gyr in the age of the old population do not change our model predictions significantly because all colors are measured with respect to the RS. Fig. \ref{Fig:logagey_Dcolour} shows the predictions for color differences in $(u-r)$ and $(g-r)$ as a function of age, for a few simple SFHs. These curves present two prominent bumps, at $\sim 10^{7}$ yr (due to the supergiant phase) and $\sim 10^{9}$ yr (due to the helium flash). Larger mass fraction $\alpha$ causes the two-burst model to approach the predictions of an SSP; as $\alpha \rightarrow 0$ the color difference curves approach zero. Two-burst models converge to color difference $\Delta = 0$ at $t = t_o$; on the other hand, the color shift of an exponentially declining SFH is not exactly zero at $t = t_o$, because $\Delta$ is measured relative to a SSP.

The age--$\Delta$ relations in Fig. \ref{Fig:logagey_Dcolour} are degenerate\footnote{In other words, age is not uniquely related to color difference.} in the range $t \lesssim 10^7$ yr, because of the supergiant ``spike"; this has little effect on our calculations because SNe\ Ia arise from stars older than $\sim 100$ Myr (\S 1), and also because the probability of observing a burst this young is very small ($\sim 10^7 \mathrm{yr} / 10^{10} \mathrm{yr} = 10^{-3}$). There is also a possible degeneracy in the age--color relation at 1.1 Gyr, due to the effects of the He flash; however this is significant only for large bursts ($\alpha \gta 0.3$), which are rare. In any case, the color and supernova rate predictions from these models are consistent with the predictions from more complex and realistic models, as we shall see.

\subsection{Age--Supernova Rate  Relation}
\label {sec:agesnr}

The formation of CO WDs starts after $t_{\mathrm{WD}} \simeq 100$ Myr (\S 1), and so  DTD$(t<t_{\mathrm{WD}}) \equiv  0$. There is in fact excellent observational evidence that SNe\ Ia do not form at early times \citep{Schawinski2009_sn, Anderson2015_young}. Whether any SN\ Ia can occur promptly after $t_{\mathrm{WD}}$ depends on the minimum merging time in the DD scenario, and on the exact treatment of mass transfer in the SD scenario.

As discussed in \S \ref{sec:intro}, a cutoff in the DTD might be expected at $t_{\mathrm{c}} \sim 10^9$ yr. We therefore model the DTD as two power laws, $\propto t^{s_1}$ for $t_{\mathrm{WD}} < t < t_{c}$, and $\propto t^{s_2}$ for $t > t_{c}$. The DTD segments are normalized so that the DTD is continuous at $t_{c}$. The normalization constant for $t < t_{c}$ is determined from
the condition DTD$(5 \times 10^{8} \textrm{yr}) \simeq 10^{-12.2} \ \textrm{SNe} \ \textrm{M}_{\odot}^{-1} \ \textrm{yr}^{-1}$ (derived from \citealt{Sullivan2006_SNrates}). (However, it should  be emphasized that the overall normalization of the DTD does not affect any of our later results on the shape of the DTD.)

We initially considered 6 cases: {\it (i)} a constant power slope $s_1=s_2=-0.5$; {\it (ii)} a soft break $s_1=-0.5, s_2=-1$; a constant power slope {\it (iii)} $s_1=s_2=-1$ and {\it (iv)} $s_1=s_2=-1.25$; {\it (v)} a break $s_1=-1, s_2=-2$; and {\it (vi)} a cutoff $s_1=-1, s_2=-3$. In all cases we assume $t_{\mathrm{WD}} = 10^8$ yr and $t_{\mathrm{c}} = 10^9$ yr; as we shall see the results are not affected by the exact values of $t_{\mathrm{WD}}$ and $t_{\mathrm{c}}$. (We denote these 6 cases ``$-0.5/-0.5$", ``$-0.5/-1$", etc..) We also considered the effects of a drastic cutoff, $s_1=-1, s_2=-\infty$, where the SN rate drops to zero past $t_c$.

The DTDs {\it (ii)} and {\it (iii)} are derived from the DD channel; the slope $s_2=-1$ for $t > t_{\mathrm{c}}$ is set by energy loss from gravitational waves (\S \ref{sec:intro}). Case {\it (i)} assumes that the behavior of the DTD remains unaltered for $t \leq t_{\mathrm{c}}$. In case {\it (ii)} we adopt the suggestion \citep{Kerkwijk2010_dtd} that the DTD at $t_{\mathrm{WD}}<t<t_{c}$ is dictated by the the formation rate of WDs, $\propto t^{-0.5}$ \citep{Pritchet2008_dtd}. Cases {\it (i)} and {\it (iv)} represent a small deviation from the DD channel prediction. Cases {\it (v)} and {\it (vi)} assume a steeper slope beyond $t_{\mathrm{c}}$, as predicted for the SD channel.  

\subsection{Color--Supernova Rate Relation}
\label{sec:color_sSNRL}

\begin{figure}
\epsscale{1.2}
\plotone{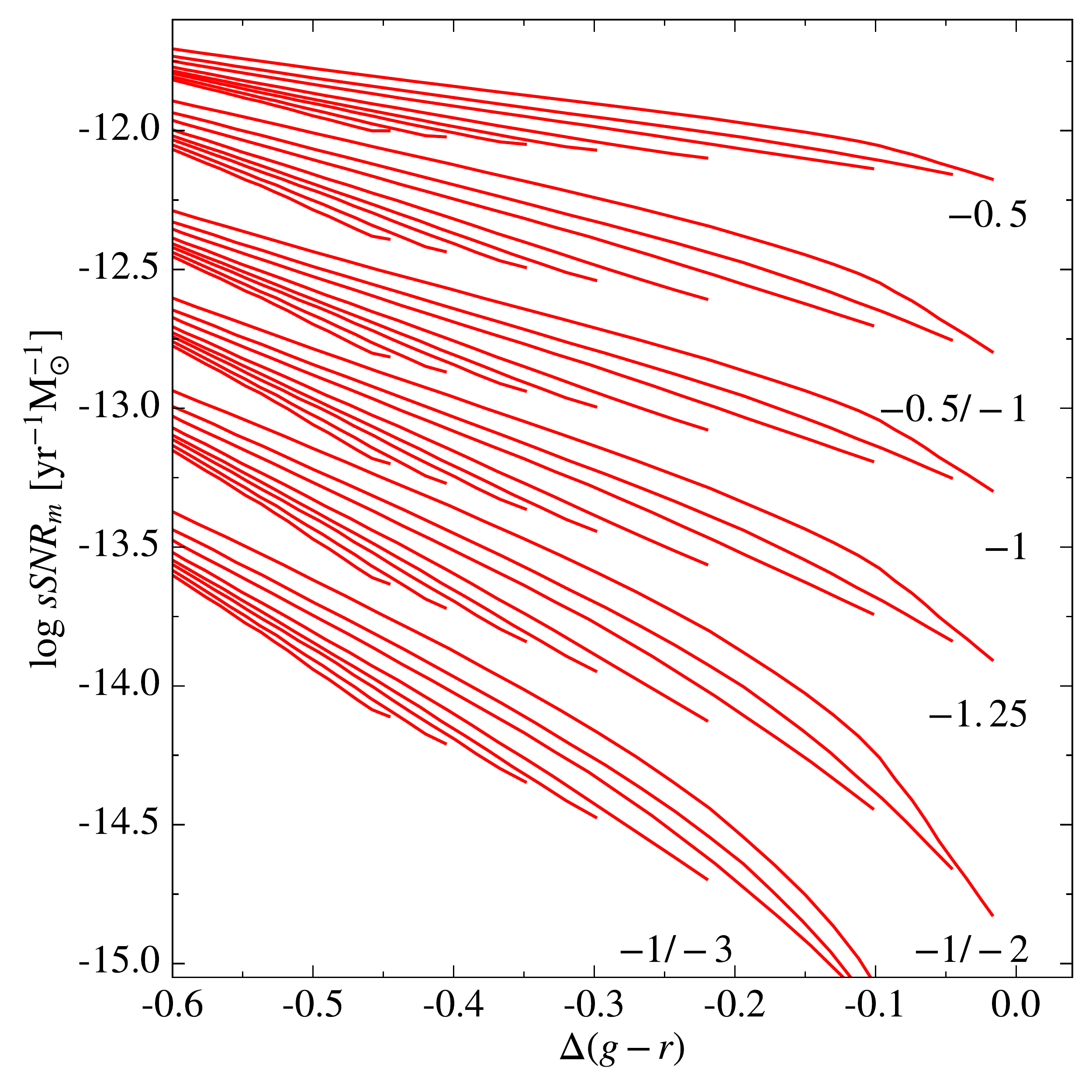}
\caption{Dependence of the SN rate per unit mass on the color shift in $g-r$ for eight exponential SFHs. For a given DTD, each curve shows the SN rate per unit mass prediction according to a given SFH; age increases to the right along each line. Plotted are the results for exponential models, where $\tau=$ 1, 1.5, 2, 3, 4, 5, 7, 10 Gyr (top to bottom). Curves for a given DTD have been vertically shifted for clarity.}
\label{Fig:logagey_CsSNR}
\end{figure}

The specific SN\ Ia rate per unit mass, $sSNR_m$, is calculated from the convolution of the DTD with SFH. The predicted values are shown in Fig. \ref{Fig:logagey_CsSNR} for exponential models; age increases to the right for each curve. There is considerable dependence on SFH in these curves (a factor of $\gta 3$ at a given color for exponential SFHs, and more when other SFHs are considered), yet not much variation with DTD. It should be noted that these exponential models give good agreement with observed galaxy colors in both the $u-r$ {\it vs.} $g-r$ diagram, and also the GALEX $NUV-r$ {\it vs.} $g-r$ diagram (e.g. \citealt{Wyder2007_cmd}); these models match not only ``red and dead'' ellipticals (short timescale $\tau$), but also bluer galaxies (larger $\tau$), which populate the green valley and the blue cloud regions of the CMD.

For any SFH we can derive the mass-to-light ratio $M/L$ using the FSPS models. (Simple analytical expressions exist for the $M/L$ of the two-burst and exponential SFH cases.) We use $M/L$ measured in the $r$-band and adopt an absolute magnitude $M_{r, \odot} \ = \ +5$ to convert to solar units (though the exact value of $M_{r, \odot}$ does not affect the final analysis). Finally, we use $M/L_r$ to convert $sSNR_m$ to $sSNR_L$, the specific supernova rate per unit $r$-band solar luminosity. 



\begin{figure}
\epsscale{1.2}
\plotone{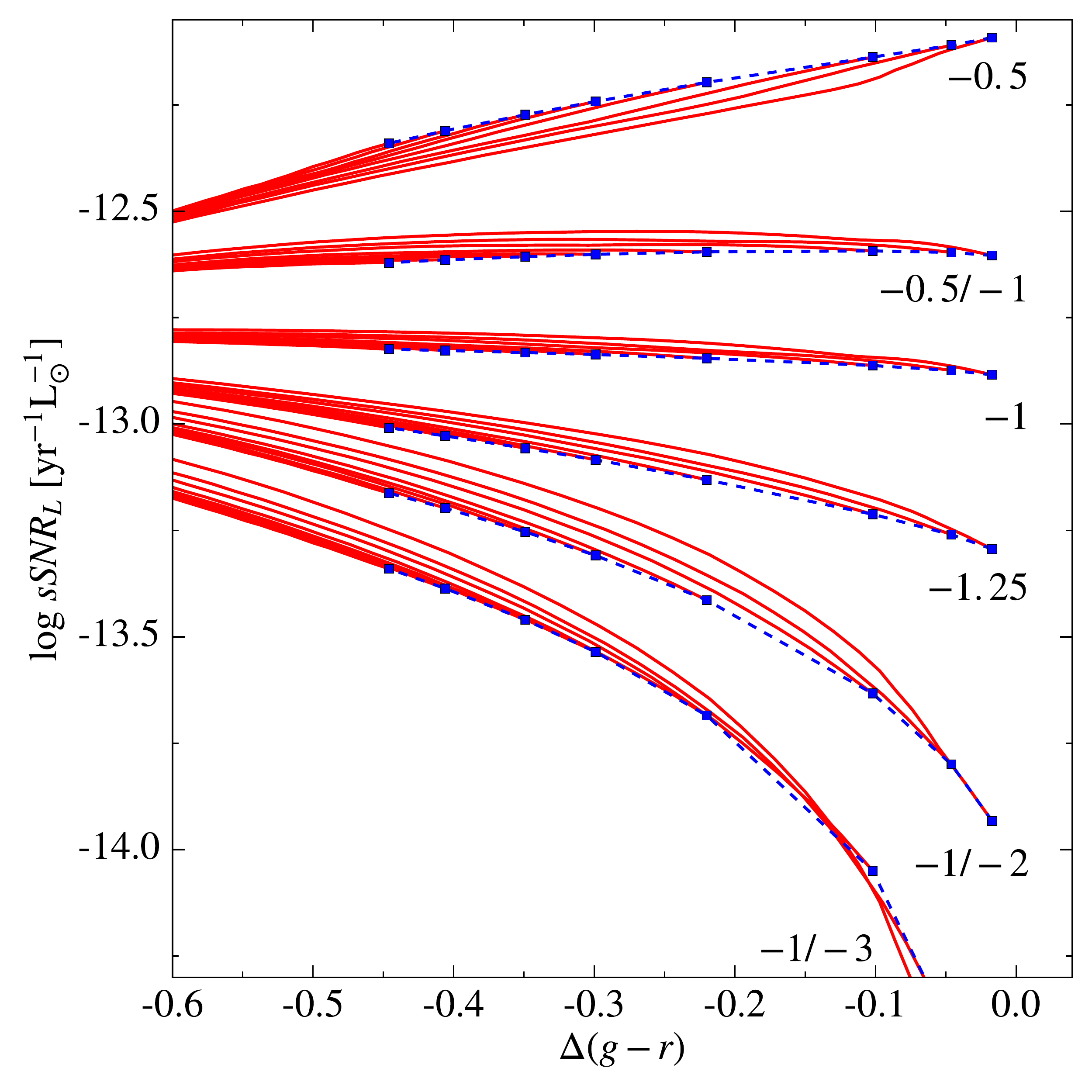}
\caption {SN rate per unit flux in the $r$-band as a function of color shift in $g-r$ (relative to the red sequence) for the same eight SFHs as in Fig. \ref{Fig:logagey_CsSNR}. Each curve spans a range of ages, with $t= 10$ Gyr on the right (blue squares), which we use to construct the $\Delta(g-r)-sSNR_L$ relation. The scatter in the predicted rate is small, implying that the SN rate per unit flux is nearly independent of the SFH. We note that the colors of the exponential models match the colors of real galaxies. Each set of DTD curves has been vertically shifted for clarity. }
\label{Fig:Dcd_SNR_gr}
\end{figure}

\begin{figure}
\epsscale{1.2}
\plotone{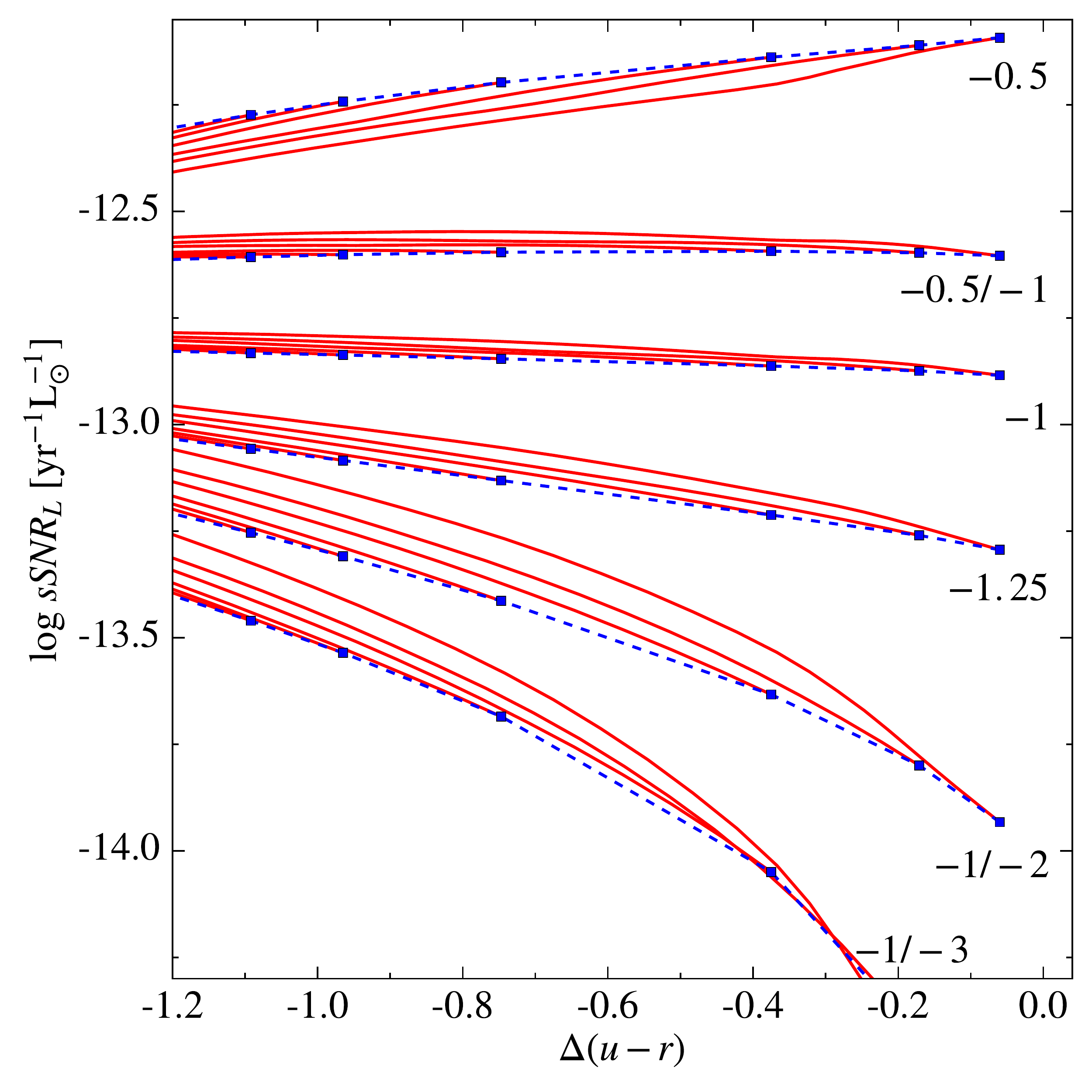}
\caption {Same as Fig. \ref{Fig:Dcd_SNR_gr} but for $u-r$.}
\label{Fig:Dcd_SNR_ur}
\end{figure}

\begin{figure}
\epsscale{1.2}
\plotone{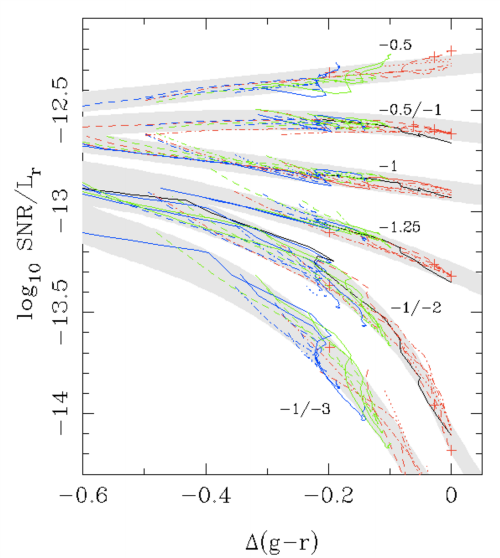}
\caption {Same as for Fig. \ref{Fig:Dcd_SNR_gr}, but with a wider range of SFH models. The curves assume distinct SFHs, which include complicated models involving infall, exponential models, and mixtures of these SFHs with pure bursts. Pure bursts and two-burst models are also included. These models do not match the colors of real galaxies as well as the exponential models, but still support the conclusion that the SN rate per unit flux is nearly independent of the SFH. Each set of DTD curves has been vertically shifted for clarity.}
\label{Fig:Dcd_SNR_SFH}
\end{figure}

Figs. \ref{Fig:Dcd_SNR_gr} and \ref{Fig:Dcd_SNR_ur} show the relation between $\Delta$ and $sSNR_L$ for $g-r$ and $u-r$. Remarkably, {\it the $\Delta - sSNR_L$ relation is only weakly sensitive to SFH}, except for the steepest cutoffs in DTD. This conclusion also holds if we consider a wider range of SFHs (double bursts, more complicated SFHs involving infall, and combinations of all of these models with bursts), as can be seen in Fig. \ref{Fig:Dcd_SNR_SFH}. Finally, the same conclusion remains valid if luminosity is derived from the $g$ or $i$ bands (because the plots are in terms of color). By way of contrast, as we have already seen, the relation between $\Delta$ and $sSNR_m$ shows a great deal of SFH-dependent scatter. 




The difference between $sSNR_m$ and $sSNR_L$ can be explained as follows. Consider an exponential burst of star formation, with SFR $\propto e^{-t/\tau}$. The spread in $sSNR_m$ at a given color (e.g. Fig. \ref{Fig:logagey_CsSNR}) is due to the difference in mean age of the stellar populations; {\it at fixed color}, populations with longer $\tau$ are older, and hence have a lower $sSNR_m$. For $\tau=1$ (10) Gyr, the mean age of stars in a stellar population that possesses $\Delta(g-r)=-0.4$ is 1.8 (5.8) Gyr. A $t^{-1}$ dependence of the DTD makes $sSNR_m$ for the $\tau=10$ Gyr population a factor of $\sim 3$ lower than for the $\tau=1$ Gyr population. But $M/L_r$ for $\tau=10$ Gyr is a factor of $3$ smaller, and so $sSNR_L = sSNR_m \times M/L$ is almost the same for the two populations. Thus the lack of dependence of $sSNR_L$ on SFH is due to a (somewhat fortuitous) cancellation of the effects of the DTD and the mass-to-light ratio. 

Another important point apparent in Figs. \ref{Fig:Dcd_SNR_gr} -- \ref{Fig:Dcd_SNR_SFH} is that the {\it shapes of the $sSNR_L$ vs. color curves are very sensitive to the DTD.} Again, this is in contrast to the $sSNR_m$ vs. color curves.





The derived $\Delta-sSNR_L$ curves are powerful because they allow an accurate prediction of the Type Ia supernova rate from two observed quantities (color and luminosity). Furthermore, these curves are robust in several ways: {\it (i)} As already pointed out, the curves in Fig. \ref{Fig:Dcd_SNR_gr}, \ref{Fig:Dcd_SNR_ur} and \ref{Fig:Dcd_SNR_SFH} are insensitive to SFH. {\it (ii)} The rate predictions for distinct DTDs are quite different, implying that observations can be used to constrain the DTD. {\it (iii)} The results are not sensitive to various parameters that affect the stellar population models (\S \ref{sec:fspstests}). This is partly due to the fact that all colors are measured differentially relative to the RS. 

\subsection {Tests of the Stellar Population Models}
\label{sec:fspstests}

To check the results from FSPS, we used PEGASE.2 \citep{Fioc1997_pegase, Fioc1999_pegase} models to produce age--color and age--luminosity relations for simple bursts. (PEGASE models use the older Padova evolutionary tracks coupled with the BaSeL \citep{Lejeune1997_BaSelI, Lejeune1998_BaSelII} spectral library.) The results were indistinguishable from FSPS, even for the GALEX NUV bandpass (which we do not use in this paper). 

We verified that our model is not strongly dependent on metallicity
$(0.006 < Z < 0.030)$ using two methods. First we checked the
dependence of the $\Delta$ vs. $sSNR_L$ relations on metallicity. While
higher metallicities result in higher SN rates, the net effect of
varying the metallicity is equivalent to a small vertical shift in
$sSNR_L$ at all $\Delta$. Such a normalization change has no effect on
our results.

In the second metallicity test we simulated the effects of the CMR on 
our results (assuming that the CMR is driven solely by metallicity).
We allowed for a variation between $Z=0.030$ for the brightest galaxies ($M_r < -21.5$) and
$Z=0.005$ for the faintest galaxies ($M_r > -18.5$). This has little effect on
the predicted color distributions, for 3 reasons: {\it (i)} SN rates change by
only $\sim 26 \%$ between the high and low
metallicity models; {\it (ii)} very few galaxies are present in the brightest and 
faintest bins, thus
diminishing their relative weight; and {\it (iii)} perhaps most importantly,
the effects of metallicity differences are mostly removed by measuring
colors differentially with respect to the RS.


We also verified that our model is not strongly dependent on the age of the old population ($6 < t_o < 12$ Gyr), the fraction of stars in the EHB $(0.0 < \textrm{EHB} < 0.5)$, or the IMF adopted. (The EHB tests assumed a fixed EHB fraction in all galaxies. We will discuss the effects of varying the age of the cutoff in the DTD, $t_{\mathrm{c}}$, in \S\ref{sec:results}.) For the representative scenario of a two-burst with $\alpha=0.01$ and $-1/-1$ DTD, the net effect of changing each parameter, while keeping others invariant, can be very well approximated by an offset in the (nearly constant) $sSNR_L$.

We have assumed that the age of the youngest SN\ Ia progenitors is 100 Myr, and that white dwarfs in the age range $40-100$ Myr do not explode (\S 1). If on  the other hand white dwarfs in this age/mass range can explode as SNe\ Ia, the late time slope of the most likely DTDs quoted in \S \ref{sec:results} become shallower by $\lesssim 0.1$.

  
\section {Comparison of Models with Observations}
\label{sec:methodology}

The goal of this paper is to use the colors of galaxies to predict
supernova rates for different stellar population mixes and delay time
distributions. These rates can be directly compared with observations
to test the validity of various DTDs, and in principle discriminate
among different SN\ Ia progenitors.

The predicted host color distribution of supernovae is calculated as follows. For each galaxy in the control sample, we convert its color  (relative to the RS), $\Delta_{i}$, to a SN\ Ia rate per $r$-band solar luminosity, $sSNR_L (\Delta_{i})$, via the relations presented in \S \ref{sec:color_sSNRL}. We then calculate an absolute SN\ Ia rate, $R(\Delta_{i})$, by multiplying the $sSNR_{L}$ by the luminosity, $R(\Delta_{i}) = sSNR_L (\Delta_{i}) \cdot f_{i}$. The flux $f_{i}$ is computed using absolute $k$-corrected $r$ magnitudes, so that it is consistent with the fluxes used to derive the M/L relations.

Generally the scatter in the models grows as one moves to bluer colors. We therefore impose a color cut $\Delta(g-r) > -0.4$, and $\Delta(u-r) > -1$ for the SN hosts and control sample; these colors correspond roughly to the color of an SSP at age 10$^9$ yr. While this indicates that our model is more sensitive to the slope of the DTD for $t > t_{\mathrm{c}}$, our SFHs also probe earlier ages, as shown in Fig. 1. 
We have verified that the exact value of this cut, which includes
part of the blue cloud, does not affect the results. The red limit is set to be $+2 \sigma$, where $\sigma$ is the standard deviation of the RS CMR. The accepted $\Delta(g-r)$ ranges of the SDSS and MENeaCS samples are similar, because the measured $\sigma$'s are nearly the same (see Table \ref{tb:RS}). We assume $sSNR_L(\Delta \geq \Delta_{max}) = sSNR_L(\Delta_{max})$, where $\Delta_{max}$ is the color at 10 Gyr of the exponential SFH model with 1 Gyr timescale.

The models and observations are compared statistically using an approach developed by \citet{Maoz2010_powerlaw} (see also \citealt{Gao2012_SNmatching, Brandt2014_bayesian}).
The probability of a galaxy hosting $n$ supernovae is given by a Poisson distribution:

\begin{equation}
L = \prod_i (\lambda_{i})^{n_{i}} e^{-\lambda_i}/n_i!,
\label{eq:likelihood}
\end{equation}

\noindent where $\lambda_i$ is the normalized predicted rate and $n_i$ is the number of SNe observed  (0 or 1) in a given galaxy. Here 

\begin{equation}
\lambda_i = A \times R_i,
\label{eq:normrate}
\end{equation}

\noindent where $A = N_{\mathrm{obs}}/\sum_i R_i$ is the normalization constant to predict the correct number of supernovae, and $N_{\mathrm{obs}}$ is the number of SNe observed in a given sample. It can be shown (e.g. \citealt{Gao2012_SNmatching}) that the logarithm of the likelihood is simply:

\begin{equation}
ln \ L = -N_{\mathrm{obs}} + \sum_{j} ln \lambda_j,
\label{eq:lnL}
\end{equation}

\noindent where $j$ runs over galaxies that hosted a SN. We use Bayesian inference with uniform priors to calculate the most likely DTD for each sample.

For simplicity, we ignore the effects of marginalizing our rate predictions over the uncertainty in $\Delta(color)$ and $r$-band magnitude. This is justified where the power-law slope is shallow, because the supernova rates are nearly constant (see Figs. \ref{Fig:Dcd_SNR_gr} and \ref{Fig:Dcd_SNR_ur}) and the $r$-band error is typically small ($\pm 0.017$). On the other hand, steeper power-laws exhibit larger scatter around the $\Delta-sSNR_L$ relation and therefore the uncertainty of these models is most likely dominated by systematic errors, which we do not take into account (a caveat of this work). The confidence levels are computed using the likelihood derived from Eq. \ref{eq:lnL}. Our results use a $68\%$ confidence level, unless otherwise stated.

The overall normalization of the masses and rates has
no effect on the shape of the DTD. Nevertheless we note in passing that our model normalization (\S \ref{sec:agesnr}) overpredicts the total number of SNe\ Ia in the SDSS sample by a factor of $\sim 2$.

\section{Results}
\label{sec:results}


We present the supernova host color distributions for the MENeaCS cluster sample in $g-r$ (Fig. \ref{Fig:MENeaCS_gr}), and for the SDSS Stripe 82 SN Survey in $g-r$ (Fig. \ref{Fig:SDSS_gr}) and $u-r$ (Fig. \ref{Fig:SDSS_ur}). All results are shown as normalized cumulative rates as a function of $\Delta (color)$, i.e. color relative to the red sequence, as computed in \S \ref{sec:rsfit}. Predictions from our simple model are also shown in these figures, for various DTDs. The predictions are smoothed with a Gaussian kernel whose standard deviation equals the $\sigma$ of the RS. The supernova host observations are smoothed by the observational uncertainty in color. Finally in Fig. \ref{Fig:contours} we show the probability for various combinations of DTD slopes $s_1$ and $s_2$; from this probability calculation we compute a statistic to compare the observed and modelled color distributions for 3 cases:{\it (i)} a continuous power-law, $s_1=s_2$; {\it (ii)} $s_1=-1.0$ (the value expected theoretically for the DTD, and observed by several authors); and {\it (iii)} $s_1 = -0.5$ (corresponding to the formation rate of WDs). The results are given in Table \ref{tb:results}. 

{\renewcommand{\arraystretch}{2}
\begin{table}[ph]
\begin{center}
\caption{Late time slope $s_2$. \label{tb:results}}
\begin{tabular}{cccc}
\tableline\tableline
Sample & $s_1 = s_2$ & $s_1 = -1$ & $s_1 = -0.5$ \\ 
\tableline\tableline
MENeaCS & $-1.26 ^{+0.57} _{-0.33}$ & $-1.35 ^{+0.66} _{-0.47}$ & $-1.49 ^{+0.71} _{-0.49}$ \\
SDSS $(g-r)$ & $-1.50 ^{+0.19} _{-0.15}$ & $-1.69 ^{+0.26} _{-0.24}$ & $-1.84 ^{+0.28} _{-0.25}$ \\
SDSS $(u-r)$ & $-1.31 ^{+0.46} _{-0.28}$ & $-1.41 ^{+0.55} _{-0.40}$ & $-1.54 ^{+0.58} _{-0.42}$ \\
\tableline
\end{tabular}
\end{center}
\end{table}


A general conclusion is that a steep cutoff ($-1/-3$ or steeper) cannot explain any of the observations.  This drastic DTD is ruled out at the $>99.99\%$ level for all samples; there are simply too many supernovae in the oldest red sequence galaxies to allow such a sharp cutoff in the DTD (Figs. \ref{Fig:MENeaCS_gr}--\ref{Fig:SDSS_ur}). Nor is a color shift observed between the RS population as a whole and the colors of the oldest SN hosts; such a shift would have been expected if SNe\ Ia in RS galaxies were produced primarily by recent bursts of star formation. We therefore conclude that at least some SNe\ Ia must occur in old stellar populations. This result is complementary to those from Graham et al. (2012), who derived the rate of SN II in galaxy clusters from MENeaCS data, and used that measurement to evaluate the star formation rate in cluster ellipticals. They found that only a small fraction of cluster SNe Ia may originate from this 'frosting' of young stars and experience a short delay time, and that cluster SN Ia rates remain an appropriate tool for deriving the late-time DTD.




\begin{figure}
\epsscale{1.2}
\plotone{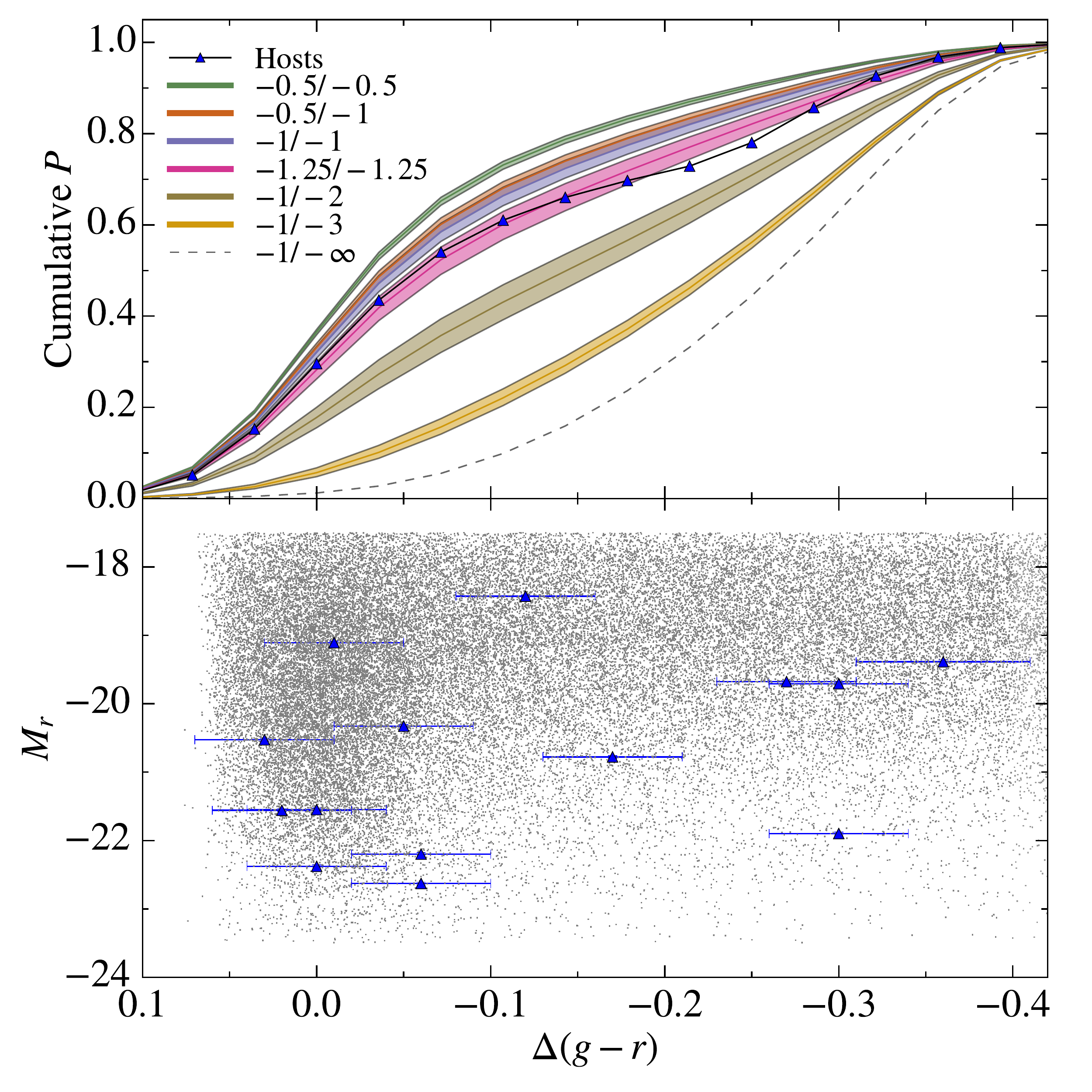}
\caption{Comparison between the predictions of the $\Delta(g-r)$--$sSNR_L$ model and observations, for the MENeaCS data sample. The lower panel shows absolute magnitudes and colors relative to the RS for all galaxies ({\it points}), and for SN\ Ia hosts ({\it filled symbols with error bars}). The upper panel shows the cumulative color distribution of observed SN\ Ia hosts compared with the predictions of various models. The top plot is color-coded according to DTD, and the filled region spans errors of $\pm 0.1$ in the slope of each DTD. The dashed line shows the predicted color distribution for a DTD with a sharp cutoff; this distribution assumes an exponential SFH with $\tau=$1 Gyr. }
\label{Fig:MENeaCS_gr}
\end{figure}


Another general conclusion is that, if the DTD is continuous in power-law slope, then this power-law slope must not be steeper than $-1.8$ at $95\%$ confidence level (all samples). The analysis of the SDSS in $g-r$ rules out continuous slopes as shallow as $-1$ at the same confidence level; such slopes are however permitted by the MENeaCS and SDSS $u-r$ samples. 




\subsection{MENeaCS}								

Fig. \ref{Fig:MENeaCS_gr} shows the cumulative observed rate and predictions. This sample is best described by a continuous power-law with $s_1=s_2=-1.26 ^{+0.57} _{-0.33}$, or by a broken power-law with $s_1=-1$, $s_2=-1.35 ^{+0.66} _{-0.47}$. If we impose $s_1=-0.5$, then $s_2=-1.49 ^{+0.71} _{-0.49}$. The large uncertainties are due to the small number of hosts in this sample. DTDs with a break, such as $-1/-2$, underpredict the SN\ Ia rates over most of the color range considered; the observed SNe\ Ia are more concentrated in the RS than these steeper models predict. As noted above, a DTD model with a strong cutoff is ruled out at the $>99.99 \%$ confidence level. Our results are marginally in agreement with \citet{Sand2012_MENeaCS_SNsurvey}, who obtain a single power-law of $-1.62 \pm 0.54$ under the assumption that early type galaxies formed in a single burst at high redshifts, and have passively evolved since then.

\subsection{SDSS}

\begin{figure}
\epsscale{1.2}
\plotone{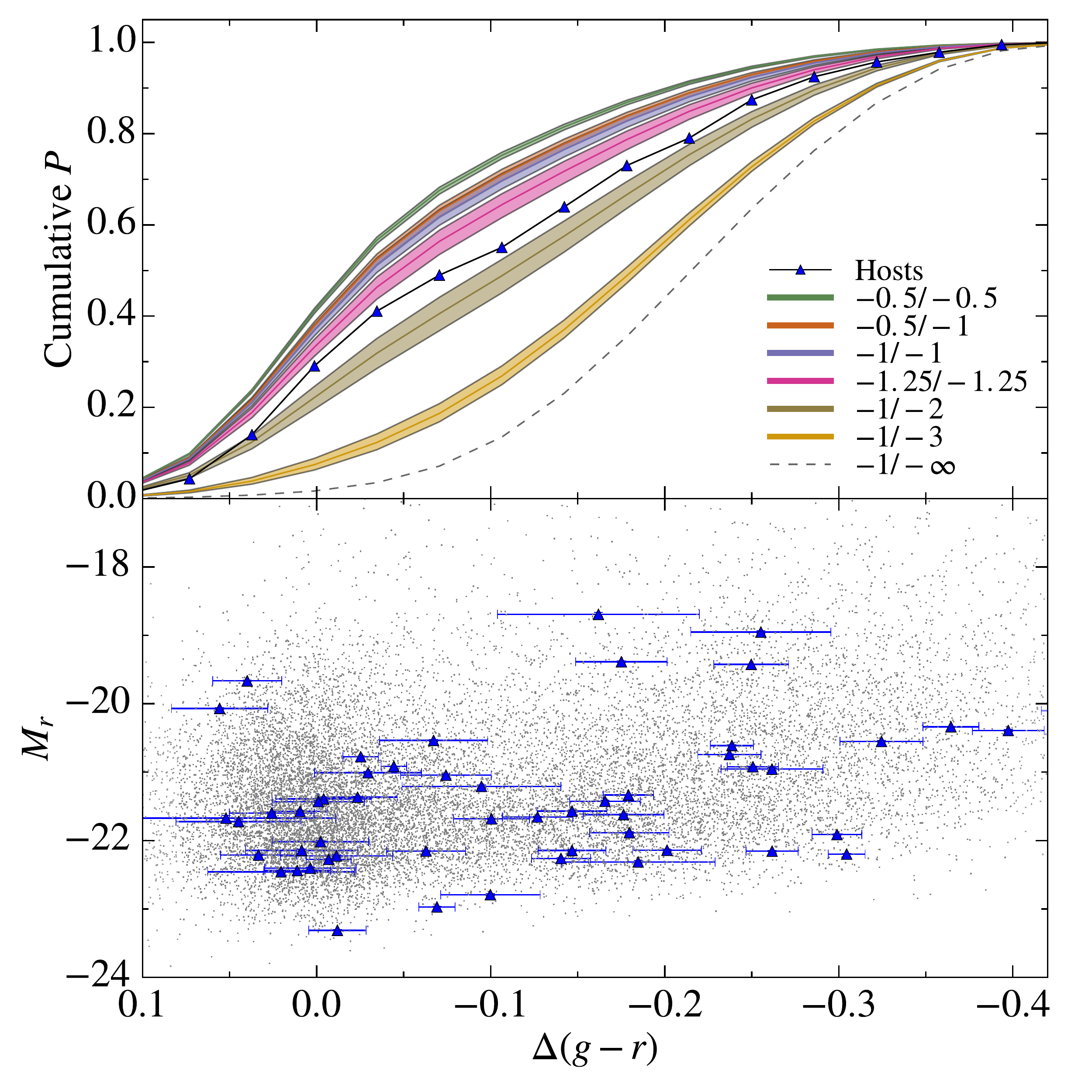}
\caption{Same as Fig. \ref{Fig:MENeaCS_gr}, but for the SDSS Stripe 82 SN sample observed in $g-r$.}
\label{Fig:SDSS_gr}
\end{figure}

\begin{figure}
\epsscale{1.2}
\plotone{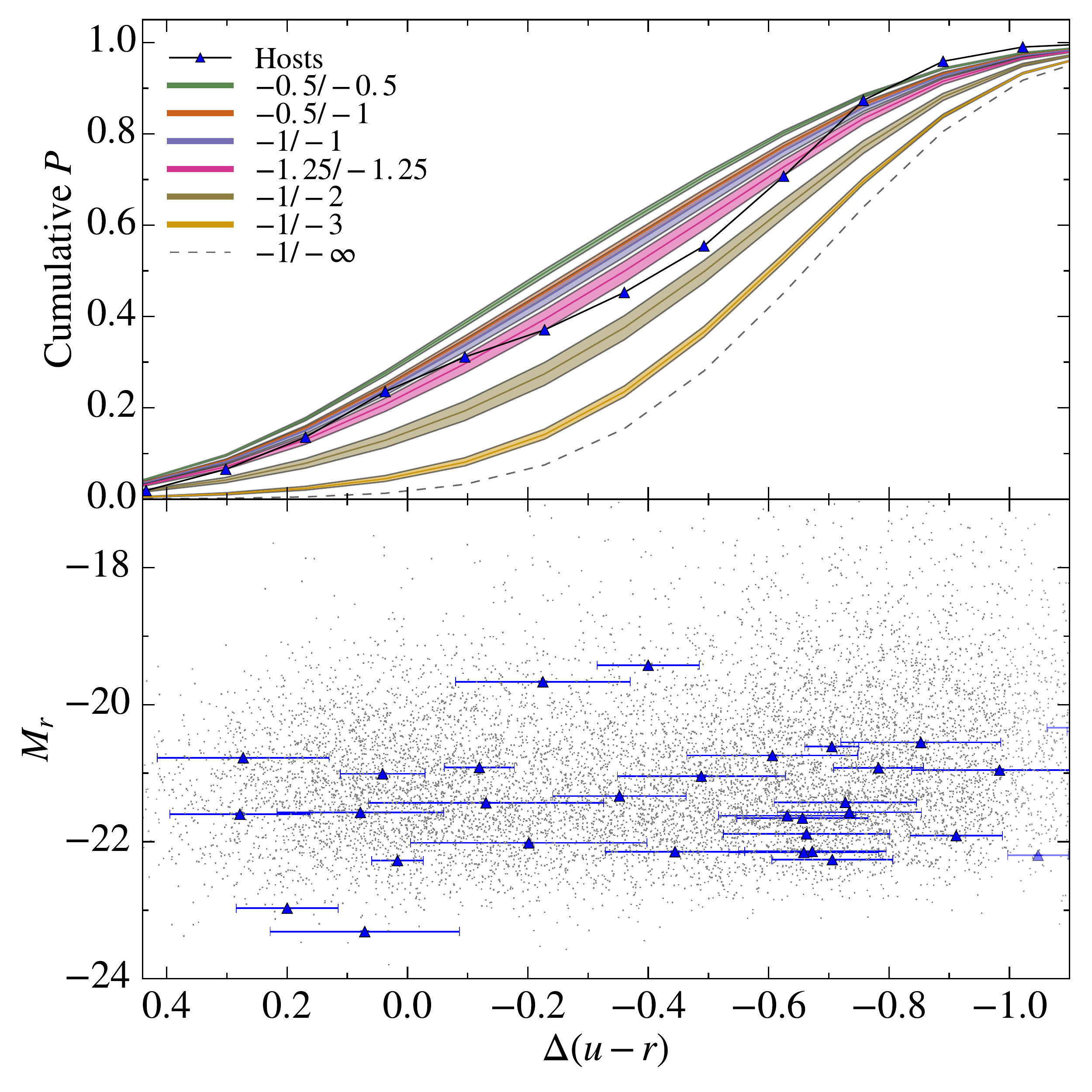}
\caption{Same as Fig. \ref{Fig:MENeaCS_gr}, but for the SDSS Stripe 82 SN sample observed in $u-r$.}
\label{Fig:SDSS_ur}
\end{figure}

The $\Delta (g-r)$ analysis is shown in Fig. \ref{Fig:SDSS_gr}. The most likely model is, for constant power-law slope, $s_1=s_2=-1.50 ^{+0.19} _{-0.15}$; alternately, if $s_1=-1$, then $s_2=-1.69 ^{+0.26} _{-0.24}$, and if $s_1 = -0.5$, then $s_2 = -1.84 ^{+0.28} _{-0.25}$. This sample, which exhibits the smallest uncertainties, suggests a slightly steeper DTD. The $s_1=-0.5/s_2=-1$ and $s_1=s_2=-1$ DTDs overpredict SNe\ Ia at the red end, causing the slope of the predicted curves to increase more steeply than the observations. The opposite trend is observed with the $-1/-2$ and $-1/-3$ models. 

The results for $\Delta (u-r)$ indicate that a somewhat shallower DTD is preferred: $s_1=s_2=-1.31 ^{+0.46} _{-0.28}$, or $s_1=-1$, $s_2=-1.41 ^{+0.55} _{-0.40}$, or $s_1=-0.5$, $s_2=-1.54 ^{+0.58} _{-0.42}$. We note that the RS is less obvious in the $u-r$ color, mainly because of the larger uncertainties in the $u$ filter. 

We have tried removing the supernovae in the faintest hosts in Figs. \ref{Fig:SDSS_gr} and \ref{Fig:SDSS_ur}. The conclusions remain unchanged.

\begin{figure}
\epsscale{1.2}
\plotone{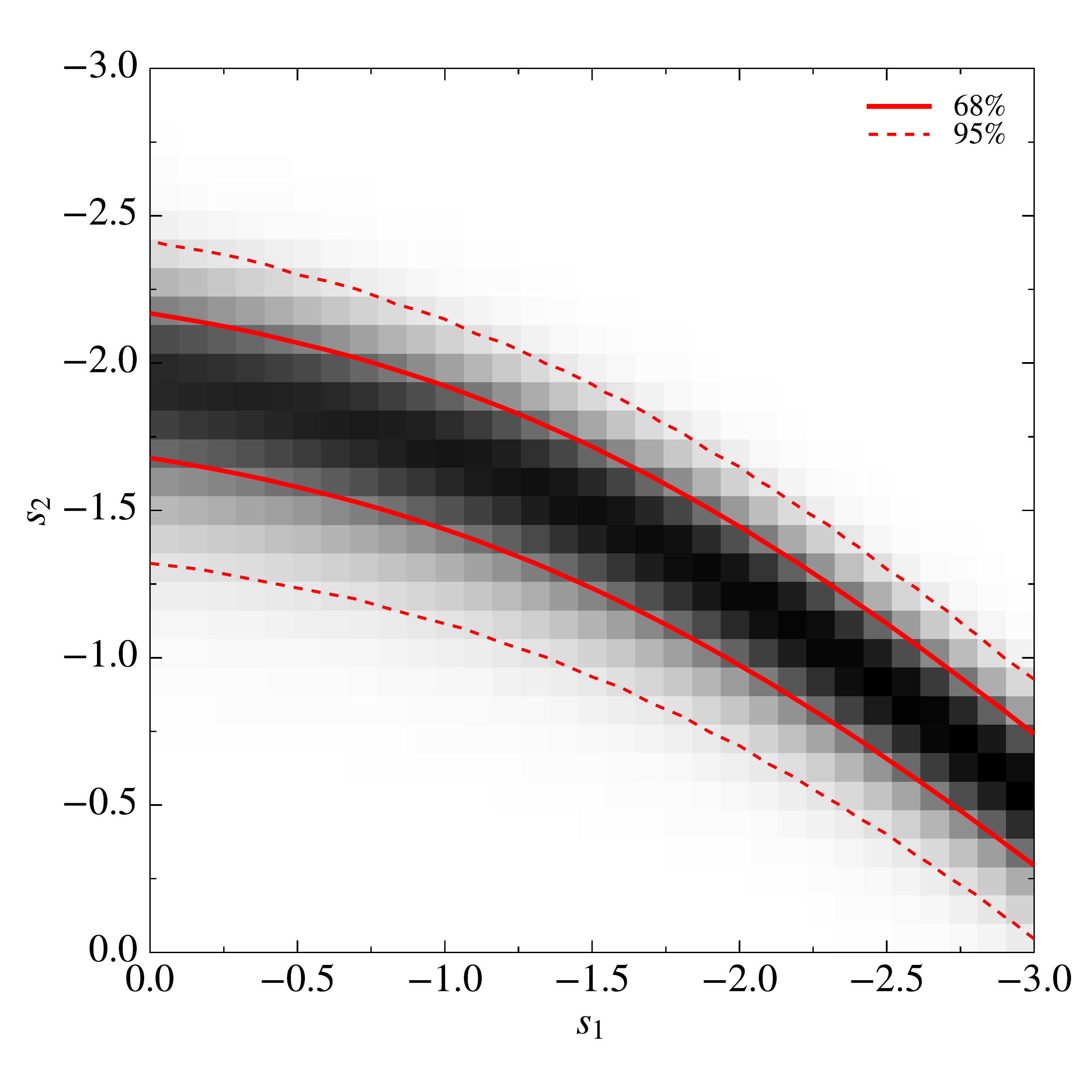}
\caption{Normalized likelihood obtained for different DTDs. The likelihoods are obtained using Bayesian statistics, where the supernova occurrence is modelled by a Poisson distribution. The DTD is parametrized as $\propto t^{s_1}$ for $t_{\mathrm{WD}} < t \leq t_{\mathrm{c}}$, and $\propto t^{s_2}$ for $ t \geq t_{\mathrm{c}}$, where $t_{\mathrm{WD}} = 10^8$ yr and $t_{\mathrm{c}}=10^9$ yr. The grid spacing is $0.1$ for both axes and the full (dashed) line shows the 68\% (95\%) contour regions. The likelihoods are computed based on the SDSS sample in $g-r$. Other data samples exhibit a similar behavior but larger uncertainties.}
\label{Fig:contours}
\end{figure}


\subsection{SDSS stretch analysis}


It has been known since the pioneering work of \citet{Phillips1993_relation} that the width of SN\ Ia 
light curves correlates with peak absolute magnitude.
The shape of SN\ Ia light curves in rest-frame $g$ or $B$ light
can in fact be matched using a simple scaling of the time axis \citep{Astier2006_SNLS}; the time scaling
factor is the ``stretch'' parameter $s$ or $x_1$. (See
\citet{Guy2010_SNLS} for the relation between these 2 quantities.) \citet{Sullivan2006_SNrates} showed that there  is a strong dependence of stretch
on host galaxy color (i.e. stellar population mix), in the
sense that redder galaxies host lower stretch SNe\ Ia.  

We obtained the stretch parameter $x_1$ for many of the SNe\ Ia in our
SDSS sample from the latest release of the SDSS-II SN survey
\citep{Sako2014_SNsurvey}, and transformed these values to $s$. We
compare the cumulative distribution of the high ($s > 1$) and low ($s
< 1$) stretch subsamples with predictions from the SDSS spectroscopic
control sample in $g-r$, as shown in Fig. \ref{Fig:SDSS_gr_s1.0}.
As we have already seen, a
single DTD, with no cutoff and power law index of $\sim -1.50 ^{+0.19} _{-0.15}$, can explain the ``All'' hosts curve.  However,
each SN\ Ia stretch sample can be represented by a distinct DTD. While low
stretch SNe are described by a continuous power law of index $-1.27 ^{+0.43} _{-0.27}$, high stretch SNe are better explained by a DTD that exhibits a break or a 
cutoff. Although a KS test between the low and high stretch SNIa host galaxies cannot rule out at a significant level that both samples were drawn from the same underlying distribution, we do find that the late-time slopes are significantly different at the 1$\sigma$ level (using the Poisson statistics in Section \ref{sec:methodology}). Specifically, for models with $s1=-1$ (i.e., the purple, magenta, brown, and yellow lines in Figure 10), we find that for the low stretch sample $s2=-1.35^{+0.51}_{-0.38}$, and for the high stretch sample $s2=-2.35^{+0.40}_{-0.41}$. We furthermore note that although the curve of the cumulative $P$ distribution for the high stretch sample in Figure \ref{Fig:SDSS_gr_s1.0} has a different shape than the other samples and models, this is not a worry because the difference is caused by a dearth of red host colors, which is fundamentally consistent with a late-time cutoff in the DTD. The interpretation of a difference in the DTD for low/high stretch SNeIa is discussed further in Section \ref{sec:discussion}.

\begin{figure}
\epsscale{1.2}
\plotone{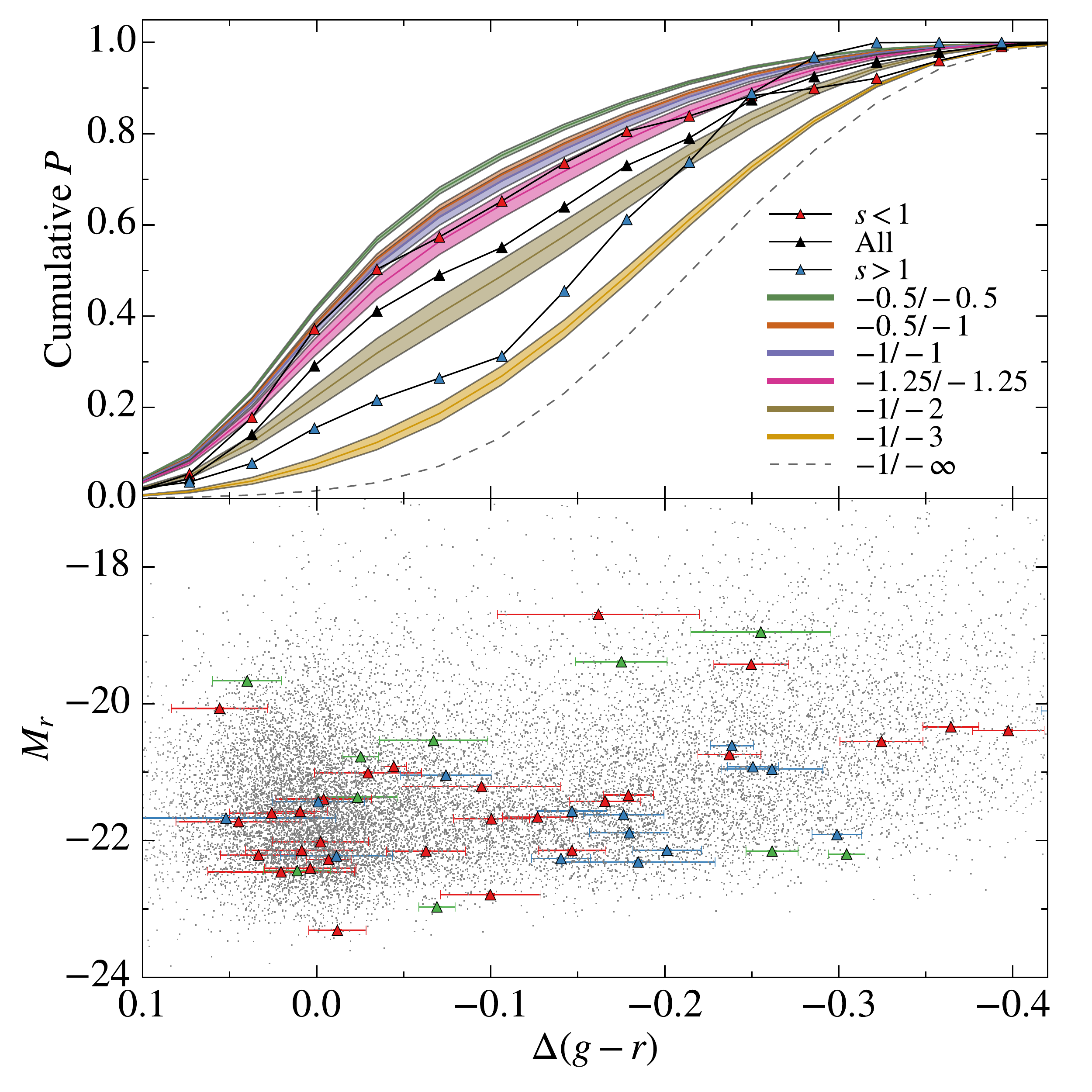}
\caption{Same as Fig. \ref{Fig:SDSS_gr}, but here the host galaxies have been subdivided according to the stretch parameter $s$ of the SNe. Low (high) stretch $s<1$ ($s>1$) objects are shown with red (blue) triangles. SNe for which the stretch was not measured are plotted in green in the CMD.}
\label{Fig:SDSS_gr_s1.0}
\end{figure}

\section{Discussion}
\label{sec:discussion}

Fig. \ref{Fig:SDSS_cutoff} shows the agreement of models and data for different cutoff slopes (using the SDSS data in $g-r$ and assuming $s_1=-1$). The most important conclusion is that DTD models with a drastic cutoff ($-1/-\infty$) can be excluded, while models with a break ($s_1 = -1$, $s_2 \leq -2.15$) are unlikely at $95\%$ confidence level.  As discussed in \S 1, this appears to be inconsistent with many formulations of the SD scenario, but more or less consistent with the DD scenario (e.g. \citealt{Greggio2005_dtd}). Increasing the cutoff age $t_{\mathrm{c}}$ adds SNe at later times (see Fig. \ref{Fig:SDSS_cutoff}), and possibly improves the agreement with the SD model. If $t_{\mathrm{c}}=2\times 10^9$, then the best model is $s_1=-1$, $s_2=-1.91 ^{+0.36} _{-0.33}$ (SDSS in $g-r$);  but DTDs with a cutoff ($-1/-3$) are still excluded at $>99.7\%$ confidence. 


\begin{figure}
\epsscale{1.2}
\plotone{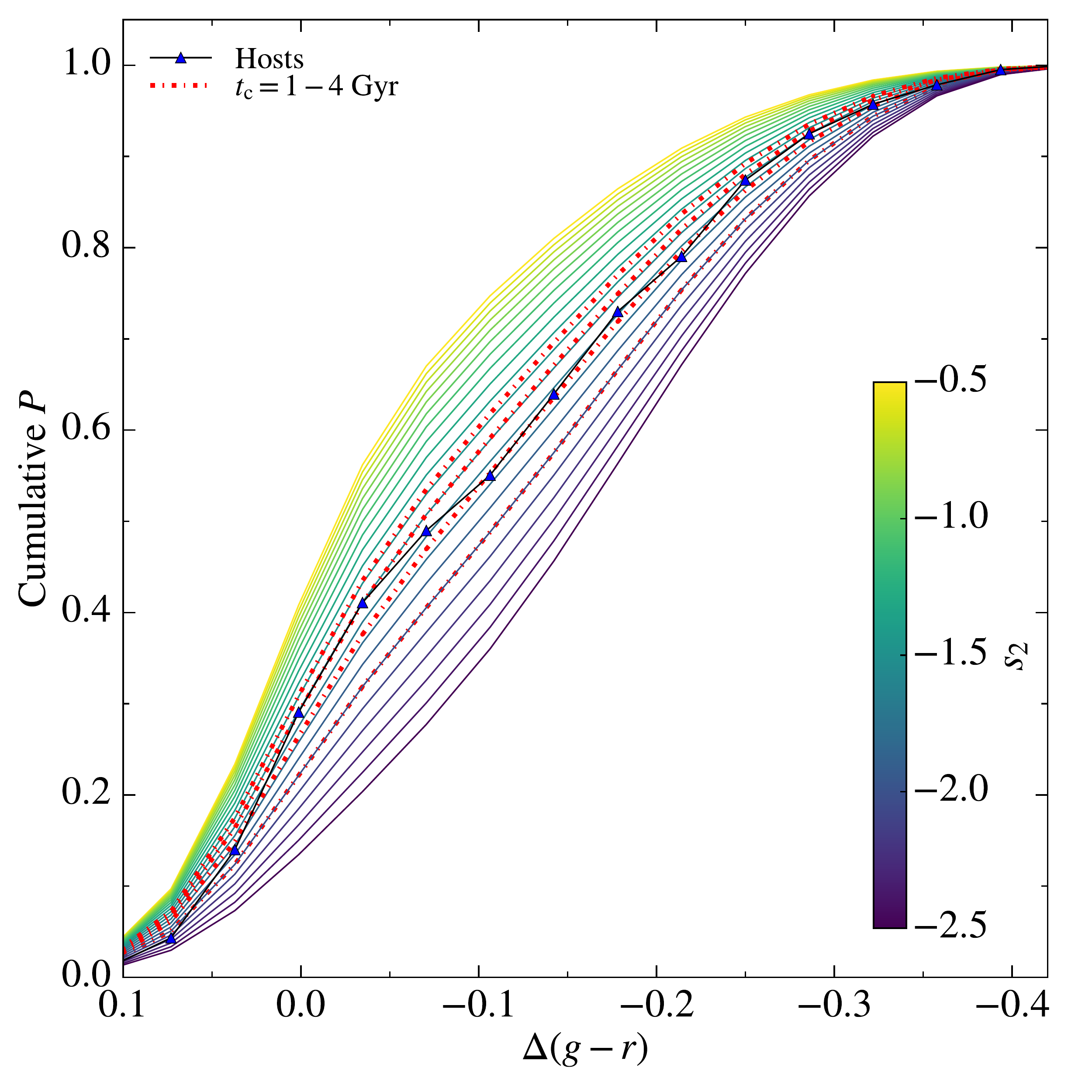}
\caption {Cumulative rate for different DTD late time slopes $s_2$ (full lines), assuming $s_1=-1$; the solid lines are for $t_{c}=10^9$ yr. The slope $s_2$ of the DTD past $t_{\mathrm{c}}$ is color coded and varies over the range $-2.5 \leq s_2 \leq -0.5$ with a step size of 0.1. Shallower models are shown in yellow (top), transitioning to steeper slopes shown in purple (bottom). The red dot-dashed lines show the effect of varying the cutoff age. The latter assumes a $-1/-2$ DTD. From bottom to top, the cutoff ages are $t_{\mathrm{c}}=$ 1, 2, 3 and 4 Gyr. These predictions were computed using the SDSS sample only, analyzed in ∆(g − r), as in Fig. \ref{Fig:SDSS_gr}.}
\label{Fig:SDSS_cutoff}
\end{figure}


The power-law slope of a continuous DTD that accounts for the rate observed in old galaxies ($t > 1$ Gyr) must be around $-1.5$ for our best sample (SDSS $g-r$); this is because of the presence of SNe\ Ia in RS galaxies in significant numbers. The other data samples (SDSS $u-r$ and MENeaCS) are consistent with this result; there are several reasons that this power-law may differ from the canonical $t^{-1}$ often quoted for the DD scenario from gravitational radiation energy losses. A DTD $\propto t^{-1}$ is derived using a distribution of initial orbital separations $\propto a^{-1}$ \citep{MaozandMannucci2012_dtd}; but this distribution of $a$ is poorly known. A bigger problem is the distribution of orbital separations emerging from the common envelope phase -- this distribution is not necessarily $a^{-1}$, nor need it be a power-law. A final problem is that the relevant time for the DTD is the sum of {\it both} the gravitational energy loss timescale {\it and} the evolutionary timescale of the secondary star.




As already discussed (\S \ref{sec:color_sSNRL}), our results are only
weakly sensitive to assumptions regarding SFH.  Nor are our
conclusions dependent on the stellar population modelling. In
\S\ref{sec:fspstests} we showed that our analysis is independent of
EHB fraction, metallicity, and age parameters; we also found similar
results with a different population synthesis code. At least part of
this insensitivity is a consequence of measuring all colors
differentially with respect to the oldest stellar populations.



Our tentative interpretation that SNe\ Ia are dominated by DD events depends on whether there is a cutoff in the DTD for the SD scenario. This cutoff is predicted from constraints on the mass of the primary (see \S \ref{sec:intro}). However, the findings of \citet{Moe2015_SD} challenge this picture by introducing a new class of binaries with extreme mass ratios, potentially leading to more SD SNe\ Ia at longer delay times. This is because the more massive star in such a binary would quickly evolve into a CO-WD, but the low mass secondary would stay in the MS phase for several Gyr before mass transfer starts. A modified DTD including these binaries is yet to be computed, and it is unclear what the expected DTD shape past 1-2 Gyr would be. Nevertheless, the possible existence of these binaries could affect our conclusions.

Subsampling by stretch parameter $s$ for the SDSS SNe, we found that
low stretch SNe\ Ia can be well represented by a DTD with a continuous power-law
slope $-1.27 ^{+0.43} _{-0.27}$, whereas the high stretch sample requires a
cutoff. A similar result has been found by \citet{Brandt2010_dtd} who employed the VErsatile SPectral Analysis (VESPA) code \citep{Tojeiro2007_VESPA} to infer the SFH of each galaxy based on its spectrum. This result was to be expected, given that
low (high) stretch SNe\ Ia are predominantly found in red (blue)
galaxies \citep{Sullivan2006_SNrates, Lampeitl2010_RSFcorrelation,
Smith2012_RSFcorrelation}.  Yet the idea of multiple SN\ Ia paths is
long-standing (e.g. \citealt{Branch1998_review}), and there is ample
independent evidence for the existence of two or more classes of SNe
Ia (e.g.  \citealt{Brandt2010_dtd, Wang2013_V0, Childress2015_Ni}).
One simple interpretation of our results is therefore that the DD
channel explains the low stretch SNe, while the SD channel may
reproduce the brighter SNe.  (See below for a discussion of other
channels.)

While the idea of two SN populations is a tantalizing explanation of
the stretch-DTD dichotomy, it should however be pointed out that this
model is not demanded by the data. As an example of a single progenitor
model, consider the effect of WD mass, which is higher in younger
(bluer) populations.  If WD mass is correlated with the brightness of
the explosion, ejecta mass and velocity, then the observed
correlations might arise naturally, without the need for two SN\ Ia
channels.


We have interpreted our results only in terms of the SD and DD scenarios; the DTDs of other channels are not as well known. For the core degenerate (CD) channel, in which a WD merges with the core of an AGB star, a SN\ Ia might result during or shortly after the common-envelope phase \citep{Livio2003_CD}. In this case, the CD scenario would be able to explain SNe\ Ia only in star forming galaxies \citep{Ilkov2013_CD}, given that the companion AGB star is required to be massive \citep{Kashi2011_CD,Livio2003_CD}. 
However, if the SN\ Ia explosion event is delayed because of the effects of rapid rotation, then the time delay necessary to spin down the object via magneto-dipole radiation would allow populations as old as 10 Gyr to host SNe\ Ia \citep{Ilkov2012_CD}.

\citet{Ruiter2011_dtd} investigated the DTD of the sub-Chandrasekhar double detonation channel and concluded that it has two distinct components. The first component exhibits short delay times ($\lesssim$500 Myr) and is derived assuming a non-degenerate He-star donor (resembling the SD scenario). The second component includes long delay times, from $\sim1-10$ Gyr, and is derived from a He WD donor (thus resembling the DD scenario). In general terms, such a DTD is consistent with our observations and models. (It is however a concern that the \citet{Ruiter2011_dtd} DTD has a power slope of $-2$ at $t\gtrsim$1 Gyr, which only marginally agrees with our results.)


\section{Summary}

The specific supernova type Ia rate per unit $r$-band luminosity, $sSNR_L$, exhibits a host galaxy color dependence that is strongly sensitive to the assumed SN\ Ia DTD, but that is insensitive to star formation history. This result occurs because of a fine balance between age, color and mass-to-light ratio effects. For a power-law DTD$\propto t^{-1}$ (as expected for the DD channel), $sSNR_L$ is nearly independent of both galaxy color and stellar population mixture. 

We have compared observations of the color distributions of SN\ Ia hosts with the predictions of simple models. We conclude that a continuous DTD with a power-law slope in the range $-1.3$ to $-1.7$ provides a good match between models and observations; a strong cutoff in the DTD for $t \gta 1$ Gyr is excluded, implying that a weak ``frosting" of young stars is insufficient to account for the comparatively large SN rate in the RS. This result is supportive of the DD channel for SN\ Ia progenitors although our most likely models deviate from the canonical power-law t$^{-1}$ predicted for this channel. We note, however, that this prediction  depends on poorly known quantities, such as the distribution of orbital separations emerging from common envelope phase.

Low stretch (fast) supernovae are better fitted by a DTD with a continuous power-law slope $-1.27 ^{+0.43} _{-0.27}$, whereas high stretch (slow) SNe\ Ia more closely match the predictions of a DTD with a break or a cutoff. This result \textit{may} be indicative of two progenitor classes; other interpretations are however possible.

Larger control and host samples will allow a better contrast between observations and the predictions of different delay time distributions. For example, relaxing the spectroscopic redshift requirement for the SDSS samples would increase the number of supernovae by an order of magnitude.  

Further modelling is needed if we are to compare observations with a wider range of progenitor scenarios; in particular, we lack predicted DTDs for the CD and sub-Chandrasekhar models. Moreover, as already noted in \S 1, the plethora of DTDs predicted for the SD channel are in strong disagreement with each other. We again highlight the possible effects of high mass ratio binaries on the shape of the SD DTD. In light of these and other complications, the interpretation of our results remains uncertain: we have good constraints on the power-law slope and (lack of a) cutoff for the DTD, but the meaning of these results in terms of progenitor scenarios is less clear.

Finally, we plan to investigate the applicability of the color--$sSNR_L$ relation to more complex SFHs, and include other parameters in its formulation. This would allow one to utilize the full color range of SN\ Ia hosts, consequently enlarging the control and host samples.

\bigskip
\bigskip

We thank Jo Bovy and Timothy D. Brandt for insightful discussions on statistical modelling of our results. CP acknowledges financial support from the Natural Sciences and Engineering Research Council of Canada.

D.J.S. is supported by NSF grants AST-1412504 and AST-1517649

Funding for the SDSS and SDSS-II has been provided by the Alfred P. Sloan Foundation, the Participating Institutions, the National Science Foundation, the U.S. Department of Energy, the National Aeronautics and Space Administration, the Japanese Monbukagakusho, the Max Planck Society, and the Higher Education Funding Council for England. The SDSS Web Site is http://www.sdss.org/.

The SDSS is managed by the Astrophysical Research Consortium for the Participating Institutions. The Participating Institutions are the American Museum of Natural History, Astrophysical Institute Potsdam, University of Basel, University of Cambridge, Case Western Reserve University, University of Chicago, Drexel University, Fermilab, the Institute for Advanced Study, the Japan Participation Group, Johns Hopkins University, the Joint Institute for Nuclear Astrophysics, the Kavli Institute for Particle Astrophysics and Cosmology, the Korean Scientist Group, the Chinese Academy of Sciences (LAMOST), Los Alamos National Laboratory, the Max-Planck-Institute for Astronomy (MPIA), the Max-Planck-Institute for Astrophysics (MPA), New Mexico State University, Ohio State University, University of Pittsburgh, University of Portsmouth, Princeton University, the United States Naval Observatory, and the University of Washington.

Based in part on observations obtained with MegaPrime/MegaCam, a joint project of CFHT and CEA/IRFU, at the Canada-France-Hawaii Telescope (CFHT) which is operated by the National Research Council (NRC) of Canada, the Institut National des Science de l'Univers of the Centre National de la Recherche Scientifique (CNRS) of France, and the University of Hawaii. This work is based in part on data products produced at Terapix available at the Canadian Astronomy Data Centre as part of the Canada-France-Hawaii Telescope Legacy Survey, a collaborative project of NRC and CNRS.

\bibliographystyle{apj}
\bibliography{biblio}

\clearpage





\clearpage

\clearpage




\end{document}